\documentclass[useAMS,usenatbib,usegraphicx,pdftex]{mn2e}
\usepackage{graphicx}
\usepackage{amsmath,amssymb}
\usepackage{pslatex}
\usepackage{float}
\usepackage{bm}
\usepackage[shortcuts]{extdash}
\bibpunct{(}{)}{;}{a}{}{,}

\title[Propeller-phase ULXs in the XMM-Newton SSC]{Searching for propeller-phase ULXs in the {\it XMM-Newton} Serendipitous Source Catalogue}
\author[H. P. Earnshaw, et al.]{\parbox{\textwidth}{H. P. Earnshaw$^{1,2}$\thanks{\vspace{-5mm}E-mail: hpearn@caltech.edu}, T. P. Roberts$^{1}$, R. Sathyaprakash$^{1}$}\\
\\
\parbox{\textwidth}{
$^1$Centre for Extragalactic Astronomy, Department of Physics, Durham University, South Road, Durham, DH1 3LE, UK\\
$^2$Cahill Center for Astronomy and Astrophysics, California Institute of Technology, Pasadena, CA 91125, USA}}

\begin{document}

\date{}

\pagerange{\pageref{firstpage}--\pageref{lastpage}} \pubyear{}

\maketitle

\label{firstpage}

\begin{abstract}

We search for transient sources in a sample of ULXs from the 3XMM-DR4 release of the {\it XMM-Newton} Serendipitous Source Catalogue in order to find candidate neutron star ULXs alternating between an accreting state and the propeller regime, in which the luminosity drops dramatically. By examining their fluxes and flux upper limits, we identify five ULXs that demonstrate long-term variability of over an order of magnitude. Using {\it Chandra} and {\it Swift} data to further characterise their light curves, we find that two of these sources are detected only once and could be X-ray binaries in outburst that only briefly reach ULX luminosities. Two others are consistent with being super-Eddington accreting sources with high levels of inter-observation variability. One source, M51 ULX-4, demonstrates apparent bimodal flux behaviour that could indicate the propeller regime. It has a hard X-ray spectrum, but no significant pulsations in its timing data, although with an upper limit of 10\% of the signal pulsed at $\sim1.5$\,Hz a pulsating ULX cannot be excluded, particularly if the pulsations are transient. By simulating {\it XMM-Newton} observations of a population of pulsating ULXs, we predict that there could be approximately 200 other bimodal ULXs that have not been observed sufficiently well by {\it XMM-Newton} to be identified as transient.

\end{abstract}


\vspace{-7mm}
\section{Introduction}
\label{sec:intro}

Ultraluminous X-ray sources (ULXs) are non-nuclear X-ray point sources with X-ray luminosities $L_{\rm X} \geq 10^{39}$\,erg\,s$^{-1}$ (for a recent review, see \citealt{kaaret17}), the majority of which are thought to be stellar-mass objects undergoing super-Eddington accretion. This was definitively demonstrated to be the case for at least some ULXs with the discovery of M82~X-2, a ULX that exhibits pulsations and is thus a neutron star (NS) undergoing highly super-Eddington accretion \citep{bachetti14}. Two other NS ULXs have since been confirmed by the detection of pulsations: NGC~7793~P13 \citep{fuerst16,israel17b} and NGC~5907~ULX-1 \citep{israel17a}. The spectral properties of these pulsating ULXs (PULXs) demonstrate them to be similar to other sources in the ULX population, with a characteristic broadened disc or two-component spectrum and a turnover at $\sim5$\,keV \citep{stobbart06,gladstone09b}, although M82~X-2's turnover is higher-energy, at 14\,keV \citep{brightman16}. This suggests that many of the ULX population could be NSs, and we may be unable to distinguish them from black hole (BH) ULXs through spectral properties alone, although estimates of the population demographics can be made by considering formation scenarios \citep{middleton17}. 

While the detection of pulsations is currently the only method of determining whether a ULX is a NS, the NS ULXs discovered so far share a common feature of transience, undergoing periods in which their flux decreases by well over an order of magnitude, giving them a bimodal flux distribution \citep{dallosso15,fuerst16,tsygankov16}. It has been suggested that this could be due to a phenomenon called the `propeller effect'. This occurs when the magnetospheric radius of the NS, at which the magnetic pressure becomes dominant over the pressure of inflowing matter, becomes greater than the corotation radius of the accretion disc, at which the Keplerian rotation of the disc matches the frequency of the stellar rotation. This has the effect of stopping accretion, and may even cause material to be ejected from the system \citep{illarionov75,stella86}. Combined with the pulsation period of the NS, the presence of this phenomenon, which provides information on the radius of the magnetosphere, allows constraints to be put on the magnetic field strength. For sources of ULX luminosity a very high, potentially magnetar-like magnetic field strength is required -- for example, $\sim10^{14}$\,G for the pulsar ULX M82~X-2 \citep{tsygankov16}. 

During this propeller-mode phase, the flux of the NS undergoes a dramatic decrease as accretion is stopped, and only increases again when the NS leaves the propeller regime and resumes accreting. Over time, this leads to a bimodal flux distribution in which `on' states and `off' states can be clearly identified. Therefore, searching for transient ULX systems with such a bimodal flux distribution may be a way of identifying other NS ULX candidates independently of detecting pulsations.

Transience is uncommon in ULXs in general -- {\it Swift} monitoring of various ULXs has shown that, while they often show variability over long timescales, for the most part they remain persistently bright (e.g. \citealt{kaaret09,grise13}). However, transient ULXs have been seen before in contexts suggesting explanations other than the propeller regime. One example is M31~ULX-1, a probable low-mass X-ray binary (LMXB) that entered the ULX regime during an outburst and subsequently underwent an exponential decay in flux \citep{middleton12}. Additionally, a pair of X-ray binaries in Cen~A reach ULX luminosities when in outburst but are most often observed at $\lesssim10^{37}$\,erg\,s$^{-1}$ \citep{burke13}. Therefore, some transient ULXs may instead be `normal', mostly sub-Eddington X-ray binaries that briefly become ULXs during outbursts. 

Another intriguing class of transient objects are Be X-ray binary (BeXRB) systems, containing an X-ray pulsar and a Be companion star (an early-type star that exhibits emission lines in its spectrum; for a review of BeXRBs see \citealt{reig11}). They can undergo powerful outbursts that allow them to briefly reach ULX luminosities, increasing in luminosity by multiple orders of magnitude in the process, before gradually declining over a period lasting days to months (e.g. SMC~X-3; \citealt{tsygankov17}). However their transience at this luminosity regime is not connected with the propeller effect, these giant outbursts are rare, and they spend most of the time at far lower luminosities, either in quiescence or undergoing smaller outbursts. 

It is also possible for previously undetected sources to be observed at ULX luminosities, such as a LMXB in M83 which has maintained a fairly steady ULX luminosity since appearing \citep{soria12}, and even to observe transience over the duration of an observation, such as in the case of two eclipsing ULXs in M51 \citep{urquhart16}.

The {\it XMM-Newton} Serendipitous Source Catalogue is a good resource for ULX discovery and study, due to the wide field of view and high effective area of the telescope, and the ready availability of fluxes for detected sources in the catalogue. Additionally, its good time resolution and ability to collect a large amount of data make it ideal for searching for pulsations in observations. Use of the telescope over the past fourteen years means that many galaxies have been observed on multiple occasions, allowing the flux of individual sources to be monitored over time. In this paper we search for transient ULXs within this catalogue and discuss the transient ULX population as observed by {\it XMM-Newton}, as well as the opportunities offered by future missions such as {\it eROSITA}.

\section{A Sample Of Transient ULXs}
\label{sec:data}

We searched for transient ULXs in a new catalogue of extragalactic X-ray sources created by matching the 3XMM-DR4 release of the {\it XMM-Newton} Serendipitous Source Catalogue \citep{rosen16} with the Third Reference Catalogue of Bright Galaxies (RC3; \citealt{devaucouleurs91}) and the Catalogue of Neighbouring Galaxies (CNG; \citealt{karachentsev04}), the method for which will be described in detail in an upcoming paper (Earnshaw et al. in prep). 

We produced a list of transient ULXs by running the positions of the sources in our catalogue identified as ULXs through {\sc flix}\footnote{\vspace{-2mm}http://www.ledas.ac.uk/flix/flix\_dr5.html}, which produces an estimate of the $3\sigma$ flux upper limit for every observation of that location by {\it XMM-Newton} where the source is not detected. We defined candidate transient objects as those with at least an order of magnitude's difference between the maximum detected flux and the minimum detected flux or flux upper limit. 

\begin{table}
\caption{The five ULXs in our sample that demonstrate high levels of variability over multiple observations.} \label{tab:srcs}
\vspace{-4mm}
\begin{center}
\begin{tabular}{lcc}
  \hline
  Name & R.A. \& Dec. & $D^a$\\ 
   & (J2000) & (Mpc)\\
  \hline
  M74 ULX-2 & 01 36 36.4 $+15$ 50 36 & 9.46 \\ 
  M106 ULX-1 & 12 18 47.6 $+47$ 20 54 & 5.97 \\ 
  M51 ULX-4 & 13 29 53.3 $+47$ 10 42 & 8.55 \\ 
  NGC 6946 ULX-1 & 20 35 00.1 $+60$ 09 08 & 6.28 \\ 
  NGC 7479 ULX-1 & 23 04 57.6 $+12$ 20 28 & 28.17 \\ 
  \hline
\end{tabular}
\end{center}
\vspace{-2mm}
$^a$The distance to the host galaxy in Mpc, found by averaging the entries in the NED Redshift-Independent Distances database obtained from the tip of the red giant branch for M74, M51 and NGC~6946, from Cepheid standard candles for M106 and from SN1a standard candles for NGC~7479.
\end{table}

Our catalogue contains 12 transient ULXs by this definition. Upon inspection, we removed one source likely to be a camera artefact, two that were blended with another source nearby, and two pairs of sources which were likely the same sources erroneously assigned two source IDs in 3XMM-DR4, for which a detection of one was a non-detection of the other. What remains is a sample of five ULXs with genuine evidence of a high level of variability over multiple observations, listed in Table~\ref{tab:srcs}. We briefly describe what is currently known about their long-term variability below.

{\bf M74~ULX\=/2} was previously identified as an ultraluminous transient in \citet{soria02}, in which it was strongly detected by {\it XMM-Newton} after being below the detection limit during two {\it Chandra} observations a few months earlier, and thus possessing an X-ray luminosity below $\sim10^{37}$\,erg\,s$^{-1}$ in those observations. The galaxy was revisited 11 months later with {\it XMM-Newton} and the source was no longer detected, meaning that its flux had decreased by at least $\sim30$ times \citep{soria04}. {\bf M106~ULX\=/1} is found in the \citet{winter06} ULX catalogue (as NGC~4258~XMM1), where it was noted as a transient source. {\bf M51~ULX\=/4} is noted as a transient source in \citet{terashima04}, having approximately two orders of magnitude's difference in flux between a detection and a flux upper limit in two {\it Chandra} observations. {\bf NGC~6946~ULX\=/1} is one of the best-studied of our sample, and has previously been found to exhibit both short- and long-term variability \citep{berghea08,fridriksson08}. We recently analysed the data from this source as part of a study of Eddington Threshold objects \citep{earnshaw17}, in which we propose that it is in an intermediate regime between being an ultraluminous supersoft source and being in the soft ultraluminous regime of super-Eddington accretion (see \citealt{sutton13}). {\bf NGC~7479~ULX\=/1} is the most luminous of our sample, and has also demonstrated over an order of magnitude variability in flux across two {\it Chandra} observations \citep{sutton12}.

\begin{figure*}
	\begin{center}
		\includegraphics[width=170mm]{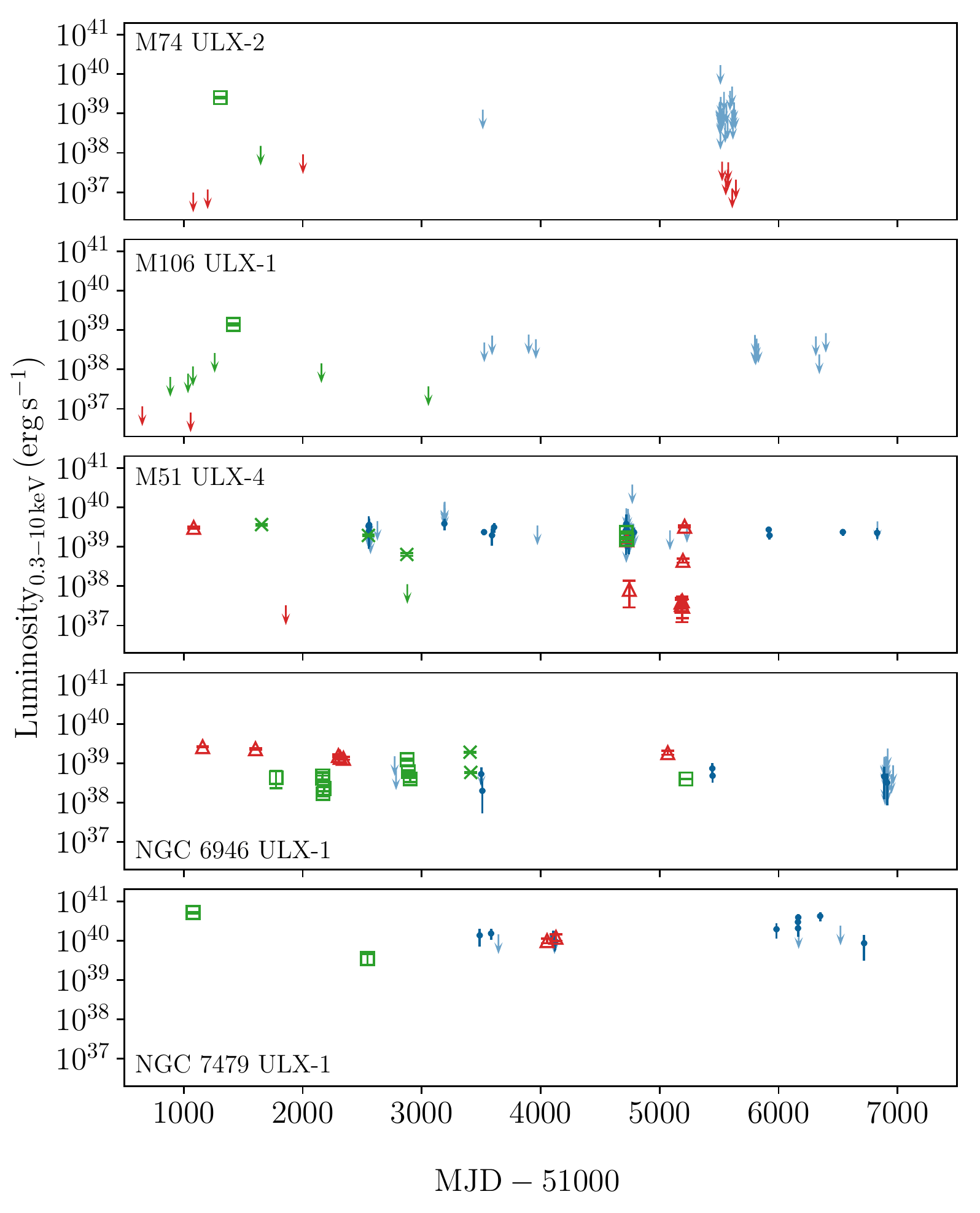}
	\end{center}
	\caption{The long-term light curves for our sample of candidate transient ULXs. {\it XMM-Newton} observations are marked in green -- clean detections are shown with square markers, and detections with an {\it XMM-Newton} quality warning flag for fluxes that could be affected by bright diffuse emission at the location of the source are marked with a cross. Upper limits calculated using {\sc flix} are shown as downward-pointing arrows. {\it Chandra} observations are marked in red, with detections shown as a triangle and upper limits as a downward-pointing arrow. {\it Swift} observations are shown in blue, with detections shown as a small circle and upper limits as a lighter-coloured downward-pointing arrow.}
	\label{fig:lc}
\end{figure*} 

We plot the {\it XMM-Newton} long-term light curves of our sample in Fig.~\ref{fig:lc}, showing {\it XMM-Newton} detections and the {\sc flix} upper limit in cases where the source was not detected in an observation of its host galaxy. We mark with crosses those detections that have a major ({\sc sum\_flag}$>1$) {\it XMM-Newton} quality warning flag associated with them, since flagged sources may be associated with bright extended emission that could contaminate the source spectrum and erroneously increase the flux calculated by the 3XMM pipeline. We also plot on the fluxes found from {\it Chandra} and {\it Swift} observations of these sources. {\it Chandra} fluxes and $3\sigma$ upper limits were found using the {\sc srcflux} routine in the CIAO software package, and {\it Swift} fluxes and $3\sigma$ upper limits were found by performing aperture photometry on the source location. In both cases, fluxes were derived from the 0.3--10\,keV count rates assuming a power-law spectrum with $N_{\rm H}=3\times10^{20}$\,cm$^{-2}$ and $\Gamma=1.7$ to make them comparable with the fluxes generated by the {\it XMM-Newton} pipeline -- we note that this spectral shape is not generally representative of ULXs, which tend on average to have softer spectra (e.g. \citealt{gladstone09b}), and some sources in this data set are already known to have far softer spectra (e.g. NGC~6946~ULX-1). 

None of the previously identified PULXs appear in our sample. We do not find NGC~5907~ULX\=/1 in our sample because {\sc flix} is currently based upon the 3XMM-DR5 dataset, and the observation in which it drops in flux \citep{walton15,israel17a} is in a later data release than 3XMM-DR5. Similarly, we do not find NGC~7793~P13 because it is not detected in 3XMM-DR4 and so does not appear in our initial list of ULXs. M82~X-2 is blended with M82~X-1 at {\it XMM-Newton}'s resolution, and so is not identified as an individual source in the {\it XMM-Newton} catalogue. 

\section{Discussion}
\label{sec:disc}

We were successfully able to use the {\it XMM-Newton} Serendipitous Source Catalogue and {\sc flix} to identify five genuinely high-amplitude X-ray variables from a catalogue of ULXs, which display a number of different variability behaviours between them.

M74~ULX-2 and M106~ULX-1 both have a single detection in {\it XMM-Newton}, and are undetected in all other observations across all instruments we consider. This behaviour shows similarity to the outbursting sources observed in M31 and Cen~A \citep{middleton13,burke13}, with a peak luminosity in the ULX regime and a quiescent luminosity below $10^{37}$\,erg\,s$^{-1}$, and so they may therefore be `ordinary' LMXBs that reach ULX luminosities during an outburst. Alternatively, they may be BeXRBs undergoing a giant outburst that reaches ULX luminosities, similar to SMC~X-3 \citep{tsygankov17} -- since the gap between the ULX detection and the subsequent upper limit extends to hundreds of days, there is sufficient time for the source to return to a lower-luminosity state even if the decay in luminosity is gradual. Brief increases in luminosity over $\sim$2 orders of magnitude into the ULX regime have also been observed in flaring sources (e.g. \citealt{sivakoff05,irwin16}), however these flares occur on the timescale of minutes, and closer examination of the light curves is required to identify whether M74~ULX-2 and M106~ULX-1 exhibit similar behaviour within their high-flux observations.

NGC~6946~ULX-1 and NGC~7479~ULX-1 show variation over an order of magnitude in luminosity, hence they are found in our sample, however it is clear from their light curves that they are persistent sources (with {\it Swift} upper limits consistent with luminosities that they have previously been observed at). NGC~6946~ULX-1 spends most of its time in the Eddington threshold luminosity regime between $10^{38}$ and $10^{39}$\,erg\,s$^{-1}$, occasionally reaching ULX luminosities, whereas NGC~7479~ULX-1 remains in the ULX regime throughout. The spectra for NGC~7479~ULX\=/1 have previously been found to be consistent with it being an extreme super-Eddington accreting stellar-mass object like the majority of the ULX population as a whole \citep{wang10,sutton12}. Similarly, while NGC~6946~ULX\=/1 is fainter and much softer, it can also be interpreted as a super-Eddington source viewed at very high inclinations \citep{earnshaw17}. This shows that super-Eddington sources can be capable of high amounts of inter-observation variability, but does not in itself provide any evidence for a NS nature of the central compact object.

\begin{figure}
	\begin{center}
		\vspace{-4mm}
		\includegraphics[width=80mm]{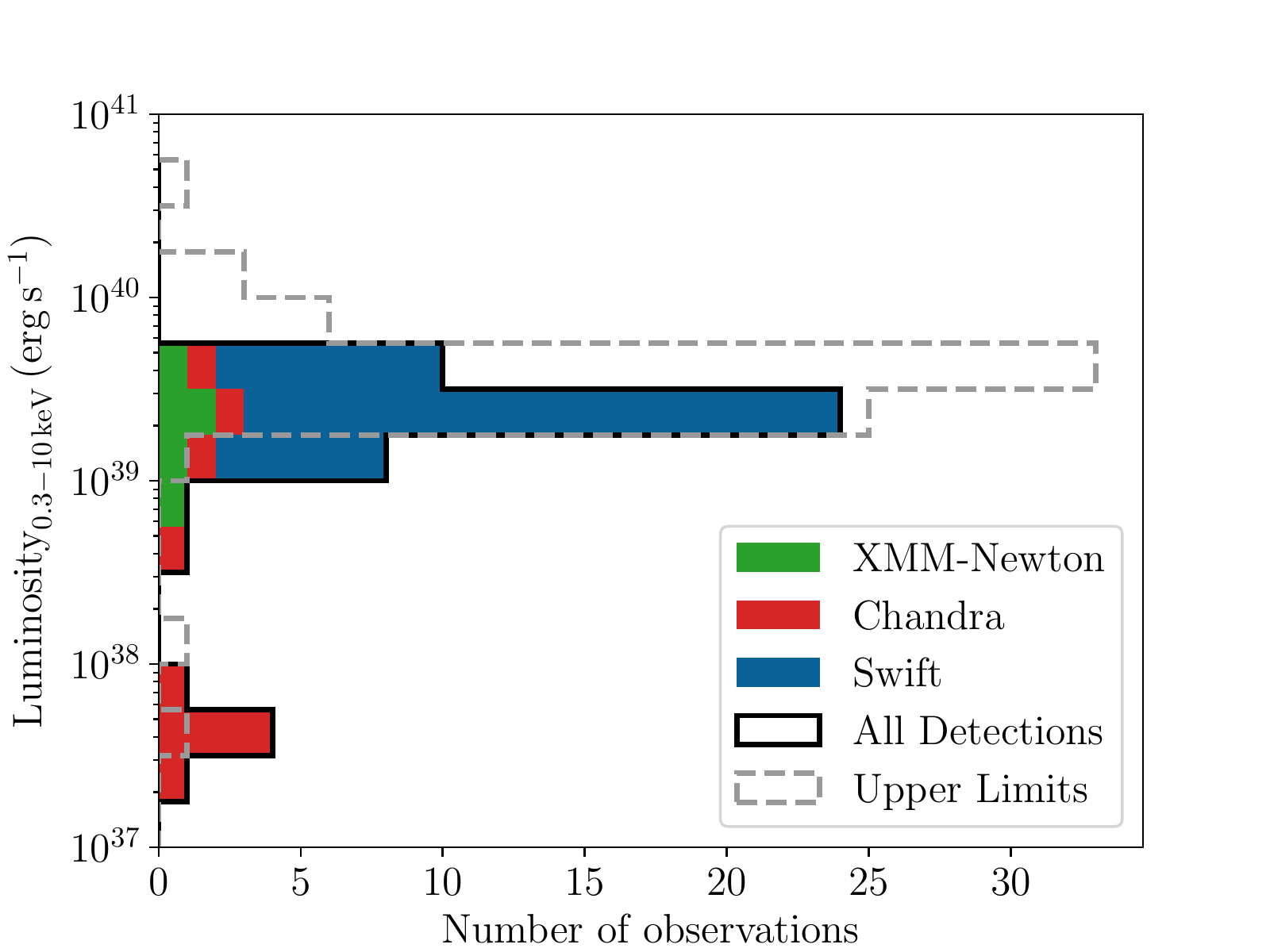}
		\vspace{-2mm}
	\end{center}
	\caption{The histogram of detected luminosities for M51~ULX-4, with bars subdivided into detections by {\it XMM-Newton} (green), {\it Chandra} (red) and {\it Swift} (blue). The distribution of upper limits is plotted with a dashed grey line.}
	\vspace{-3mm}
	\label{fig:m51hist}
\end{figure}

Of the sources in our sample, M51~ULX-4 is the only one that shows evidence of having a genuinely bimodal distribution, with a fairly consistent bright X-ray luminosity of $\sim1$--$3\times10^{39}$\,erg\,s$^{-1}$ and dim periods where the luminosity drops to $<10^{38}$\,erg\,s$^{-1}$. Single {\it XMM-Newton} and {\it Chandra} observations are the exception to this, with luminosities between $10^{38}$ and $10^{39}$\,erg\,s$^{-1}$. This could still be consistent with a generally bimodal flux distribution if the source was transitioning between accreting and propeller states during the course of those observations. We show a histogram of the observed luminosities in Fig.~\ref{fig:m51hist} which clearly illustrates the bimodal nature of M51~ULX-4's flux distribution. The two states differ for the most part by a factor $\sim40$ in flux, comparable to the ratio between high and low states exhibited by M82~X-2 \citep{tsygankov16} -- like M82~X-2, it appears that the source may not completely stop accreting in its low-flux state, which may be due to a small fraction of the surrounding matter leaking into the magnetosphere and being accreted during the propeller phase (e.g. \citealt{doroshenko14}).

We examined all available archival {\it XMM-Newton} and {\it Chandra} observations of M51~ULX-4. Its spectrum, when luminous, is ubiquitously hard, with photon indexes of $\Gamma\sim1.2$--1.7. We extracted timing data from the {\it XMM-Newton} EPIC-pn detector, which has sufficient time resolution to detect the $\sim$1\,Hz pulsations seen in PULXs, but we found no signal stronger than 1.5$\sigma$ significance in the {\it XMM-Newton} PSDs. We therefore attempted blind acceleration searches on the data, as used in other PULX detections, using the {\sc presto} software \citep{ransom01}. No pulsations were detected to greater than 3$\sigma$ significance, although we note the datasets were relatively short ($\lesssim$18\,ks). However, possible weak pulsed signals were recovered (significances of 1.5--2.6$\sigma$) in the 1.3--3.2\,Hz range, that constituted up to $\sim10\%$ of the EPIC-pn flux in these observations (upper limit to $3\sigma$ significance). This object therefore remains a plausible PULX candidate, with deeper {\it XMM-Newton} observations likely to provide the crucial diagnosis.

As part of the process of creating the ULX catalogue, we determined which of the RC3 and CNG galaxies observed by {\it XMM-Newton} were complete, i.e. the galaxies for which we are confident that all ULXs that can be detected by {\it XMM-Newton} have been (Earnshaw et al. in prep). We found there to be 441 such galaxies, including all host galaxies of our candidate transient ULX sample. Given that three sources with behaviour that could be described as transient have been detected within the sample of high-amplitude ULX variables in this complete subset of galaxies, we estimate there to be one transient ULX per $\sim1.5\times10^{13}$\,M$_{\odot}$. However this is a lower limit, since many of the ULXs in the complete sample will not have been observed a sufficient number of times with {\it XMM-Newton} for them to be identified as transient even if they are.

In order to estimate how many high-amplitude ULX variables we may be missing due to the cadence of {\it XMM-Newton} observations, we ran a simulation of 10,000 PULX-like transient sources, capable of being in two states -- a bright, accreting state with $L_{\rm X} \geq 10^{39}$\,erg\,s$^{-1}$ or a dim, propeller state with $L_{\rm X} \leq 10^{38}$\,erg\,s$^{-1}$ -- with a duty cycle of 50\% for simplicity. Each source was `observed' a number of times randomly selected from a distribution of the number of {\it XMM-Newton} visits to the RC3 and CNG galaxies, in order to simulate the frequency at which {\it XMM-Newton} observes other galaxies (this comes with the caveat that galaxies containing `interesting' sources such as ULXs are more likely than others to be observed more than once, however a majority of ULXs have still only been observed once by {\it XMM-Newton}). Only 26\% of sources were observed more than once, meaning that we do not have the ability to judge whether nearly three quarters of the sources exhibit long-term high-amplitude variability or not. For the sources with multiple observations, only those sources with at least one observation in each state were counted as transient (as otherwise they would appear as persistent sources in either the bright state or the dim state). 65\% of the sources observed multiple times were successfully identified as transient, or 17\% of the total population.

We also differentiated between two different variability scenarios -- `outbursts', in which only one bright state is observed and the source could be mistaken for a classic transient, and `bimodal', in which there are at least two observations of the source in a bright state, making the bimodal nature of its flux distribution apparent. Of the sources identified as transient, around 60\% appear as an outburst and 40\% as bimodal.

However, this does not take into account the typical depth of {\it XMM-Newton} observations, i.e. whether an observed PULX-like source is confidently detected when in the bright state, or a sufficiently low flux upper limit established when in the dim state. We therefore considered, for observations of PULXs in a bright state, the probability of any one observation of a galaxy being complete to $10^{39}$\,erg\,s$^{-1}$, meaning that we can be confident that a ULX-luminosity source within that galaxy is detected. For current {\it XMM-Newton} observations of RC3 and CNG galaxies, this probability is 31\%. For observations of PULXs in a dim state, we required the observations to be complete to $10^{38}$\,erg\,s$^{-1}$ so that a maximum source luminosity upper limit of $10^{38}$\,erg\,s$^{-1}$ could be established, the probability for which is 13\%. (The completeness of galaxy observations is dependent on the distance to the galaxy being observed and the {\it XMM-Newton} observation length, but we consider the fraction of complete observations as the probability, which is sufficient for this simulation).

When the observation depth is considered, we found that only $\sim2\%$ of the simulated PULX sample are successfully identified as transient based on {\it XMM-Newton} observations of other galaxies. Additionally, with a duty cycle of 50\%, three-quarters of these sources are observed as a single outburst, with only $\sim0.5\%$ of the entire sample observed as bimodal. Since we have observed one bimodal source, M51~ULX-4, we might therefore expect that there are up to 200 other sources with bimodal flux distributions in the {\it XMM-Newton} footprint that have not been identified as such using archival {\it XMM-Newton} data due to being insufficiently observed to reveal their transient nature. 

The ability to discover candidate PULXs in this way also depends upon the expected duty cycle of a NS entering and leaving the propeller regime. M82~X-2 has been found to be in accreting and propeller states a similar number of times over the course of 15 years of {\it Chandra} observations \citep{tsygankov16}, which suggests a duty cycle of $\sim50\%$. However, NGC~5907~ULX-1 only has one {\it XMM-Newton} observation out of five appear to drop into a propeller regime, along with a handful of {\it Swift} observations with an upper limit significantly lower than its accreting state flux \citep{israel17a}. NGC~7793~P13 appears to spend longer in the propeller regime, but is still limited to a single `episode' between 2011 and 2014, being seen at its accreting flux at all other times \citep{fuerst16,israel17b}. If the typical duty cycle of NS ULXs is significantly higher or lower than 50\%, the percentage identified as transients decreases, so that there may be a much higher number that are not identified as transients from archival {\it XMM-Newton} data. 

\begin{figure}
\begin{center}
\vspace{-4mm}
\includegraphics[width=80mm]{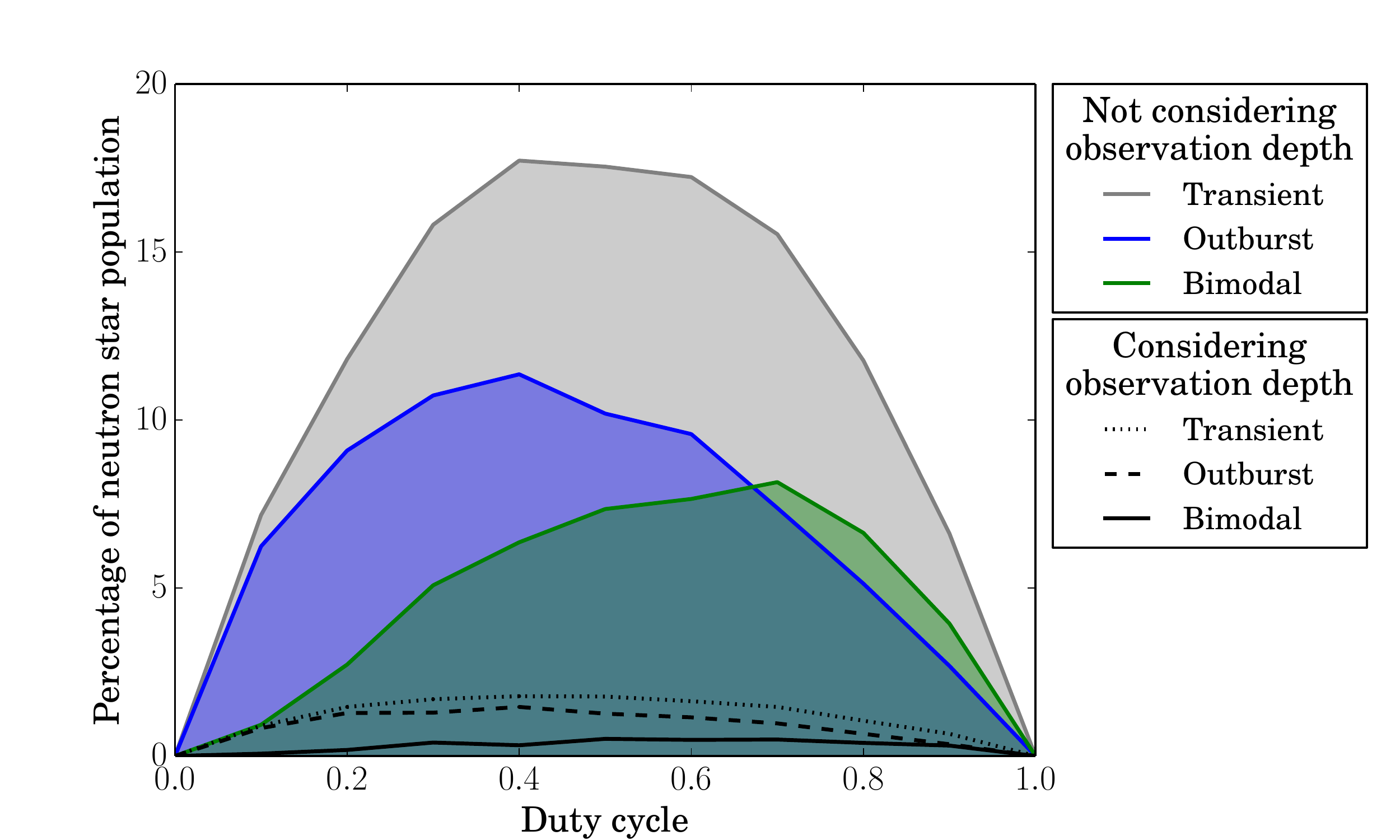}
\vspace{-2mm}
\end{center}
\caption{The percentage of the simulated population of PULX-like sources that are successfully identified as transient by {\it XMM-Newton's} current observation coverage, both before and after considering the depth of {\it XMM-Newton} observations. Sources identified as transient are further divided into `outbursts', with one single bright observation, and `bimodal' sources which have two or more bright observations.}
\vspace{-3mm}
\label{fig:dc}
\end{figure} 

In Fig.~\ref{fig:dc} we show the results of the above simulation for a range of duty cycles, both before and after the observation depth of {\it XMM-Newton} is taken into account. The total number of transient sources successfully detected and identified as such is maximised if the duty cycle of the population is 50\%, however sources with a slightly greater amount of time in a bright state are more likely to be observed with a true bimodal flux distribution, rather than as what appears to be a single outburst. In reality, the duty cycle of the accreting/propeller regimes for NSs is likely to be dependent on the conditions of each individual system.

While the cadence of {\it XMM-Newton} observations is generally insufficient for identifying the majority of bimodal-flux sources that may exist, future missions may provide a better opportunity for searching for candidate PULXs. {\it eROSITA} is the primary instrument on the upcoming X-ray satellite {\it Spectrum-R{\"o}ntgen-Gamma}, due to launch in the autumn of 2018. {\it eROSITA} is planned to perform a systematic all-sky survey with eight scans of the entire sky over the course of four years \citep{merloni12}. Using our simulation of a PULX population, eight observations is sufficient to successfully identify 96\% of all detectable bimodal sources on the sky with a duty cycle of 50\% (this decreases to $\sim50\%$ of sources for duty cycles of 20\% or 90\%). However, the average amount of observation time for each point on the sky per scan is 250\,s, which means that we can only place a upper limit of $10^{38}$\,erg\,s$^{-1}$ on non-detections for sources within 2.2\,Mpc. While this makes the vast majority of the ULX population too distant for us to perform this monitoring using {\it eROSITA}, there are 83 objects within 2.2\,Mpc between the RC3 and CNG catalogues, including all of the Local Group and some galaxies of interest such as NGC~55 and NGC~300. For these nearby galaxies and dwarfs, we are very confident that we will be able to identify a large proportion of the bright bimodal sources they contain.

The closer to the poles of {\it eROSITA's} orbit, the more exposure a source will receive. It is estimated that about 1\% of the sky will receive $>10$\,ks exposure over the course of the four-year survey, amounting to at least 1.25\,ks per scan. This would extend our ability to identify PULX candidates out to $\sim7.5$\,Mpc, but only within ten degrees of the orbital poles. Assuming that they are aligned with the celestial poles (as an illustration), this therefore only applies to galaxies with $|$Dec$|\gtrsim80$\,degrees. There is only one such galaxy within 7.5\,Mpc between the RC3 and CNG galaxy catalogues, LEDA~95597, which is not known to contain any ULXs. However, the exact position of {\it eROSITA's} orbital poles may allow a small number of other galaxies within this distance to be monitored for PULX candidates. Therefore we conclude that while {\it eROSITA} will not undertake a deep enough survey to monitor most of the ULX population (although future generation all-sky surveys may prove invaluable for this kind of monitoring), it will be capable of monitoring the X-ray populations of a reasonable number of nearby galaxies and discovering further candidate PULXs where they exist in those galaxies.

The {\it XMM-Newton} Serendipitous Source Catalogue is a useful tool for identifying transient ULXs, and the addition of archival data from other telescopes such as {\it Chandra} and {\it Swift} gives us sufficient information to more accurately characterise the long-term variability of sources to help determine their underlying nature. However, the current coverage of {\it XMM-Newton} may be wholly inadequate for detecting large numbers of transient ULXs, due to the lack of repeat observations. The planned all-sky survey by {\it eROSITA} presents an opportunity to monitor nearby galaxies across the entire sky for new PULX candidates, but will not be deep enough to monitor the majority of known ULXs. We therefore also recommend dedicated observing campaigns involving repeated, sensitive observations of galaxies containing ULXs with {\it XMM-Newton} and {\it Chandra} in order to increase the likelihood of successfully identifying PULX candidates from their long-term variability properties.

\vspace{-3mm}
\section*{Acknowledgements}

We thank the anonymous referee for helpful comments that improved this paper. We gratefully acknowledge support from the Science and Technology Facilities Council (HPE through grant ST/K501979/1, SR through ST/N50404X/1, TPR through ST/P000541/1). HPE acknowledges support under NASA contract NNG08FD60C.

This research made use of data obtained from the 3XMM {\it XMM-Newton} Serendipitous Source Catalogue, based on archival observations with {\it XMM-Newton}, an ESA science mission with instruments and contributions directly funded by ESA Member States and NASA. We also acknowledge the use of public data obtained from the {\it Chandra} Data Archive and the {\it Swift} Data Archive.

\bibliography{transientulxsletter}
\bibliographystyle{../mn2e}
\bsp

\end{document}